\documentclass[10pt]{article}

\usepackage{amsmath}
\usepackage{amssymb}
\usepackage{multirow}
\allowdisplaybreaks

\usepackage{graphicx}

\usepackage[usenames]{color} 

\usepackage{setspace} 
\usepackage{url}

\usepackage{multirow}
\usepackage{dsfont}
\usepackage{verbatim}
\usepackage{longtable}
\usepackage{esvect}
\usepackage{mathrsfs}
\usepackage[mathscr]{eucal}
\usepackage[ruled,vlined]{algorithm2e}

\usepackage{common}

\usepackage[labelfont=bf,labelsep=period,justification=raggedright]{caption}

\begin{document}

\begin{flushleft}
{\Large
\textbf{A Bayesian test to identify variance effects}
}
\\
Bianca Dumitrascu$^{1}$,
Gregory Darnell$^{1}$,
Julien Ayroles$^{1,2}$,
Barbara E Engelhardt$^{3,\ast}$
\\
\bf{1} Lewis-Sigler Institute, Princeton University, Princeton, NJ, USA
\\
\bf{2} Department of Ecology and Evolutionary Biology, Princeton University, Princeton, NJ, USA
\\
\bf{3} Department of Computer Science, Center for Statistics and Machine Learning, Princeton University, Princeton, NJ, USA
\\
$\ast$ E-mail: Corresponding bee@princeton.edu

\end{flushleft}

\begin{abstract}
\noindent Identifying genetic variants that regulate quantitative traits, or QTLs, is the primary focus of the field of statistical genetics. Most current methods are limited to identifying mean effects, or associations between genotype and the mean value of a quantitative trait.
It is possible, however, that a genetic variant may affect the variance of the quantitative trait in lieu of, or in addition to, affecting the trait mean. 
Here, we develop a general methodological approach to identifying covariates with variance effects on a quantitative trait using a Bayesian heteroskedastic linear regression model. We show that our Bayesian test for heteroskedasticity (BTH) outperforms classical tests for differences in variation across a large range of simulations drawn from scenarios common to the analysis of quantitative traits. We apply BTH to methylation QTL study data and expression QTL study data to identify variance QTLs. When compared with three tests for heteroskedasticity used in genomics, we illustrate the benefits of using our approach, including avoiding overfitting by incorporating uncertainty and flexibly identifying heteroskedastic effects. 
\end{abstract}

\section*{Introduction}

Identifying genotypic variation that impacts complex traits is central to the study of statistical genetics~\cite{WTCCC2007,stranger2007relative}.  Quantitative trait loci (QTLs) are genetic variants that are associated with differences in mean phenotype values within a population. Recently, variance QTLs (vQTLs), or genetic variants associated with differences in the variance of a quantitative trait, have been observed in genetic studies~\cite{Pare2010,yang2012fto,Brown2014,Ayroles2015,metzger2015selection}.
These studies include diverse quantitative phenotypes, including left-right turning tendency in the fruit fly \emph{Drosophila melanogaster}~\cite{Ayroles2015}, coat color in the rock pocket mice \emph{Chaetodipus intermedius}~\cite{nachman2003_rainbow_mice}, and thermotolerance~\cite{queitsch2002_hot_hsp} and flowering time~\cite{salome2011_time_Athaliana} in the plant \emph{Arabidopsis thaliana}. Robust statistical methods to identify these variance effects are essential to characterizing the role that variance effects play in the genetic regulation of complex traits including disease risk.

\begin{figure}
\centering
\includegraphics[scale=1.20]{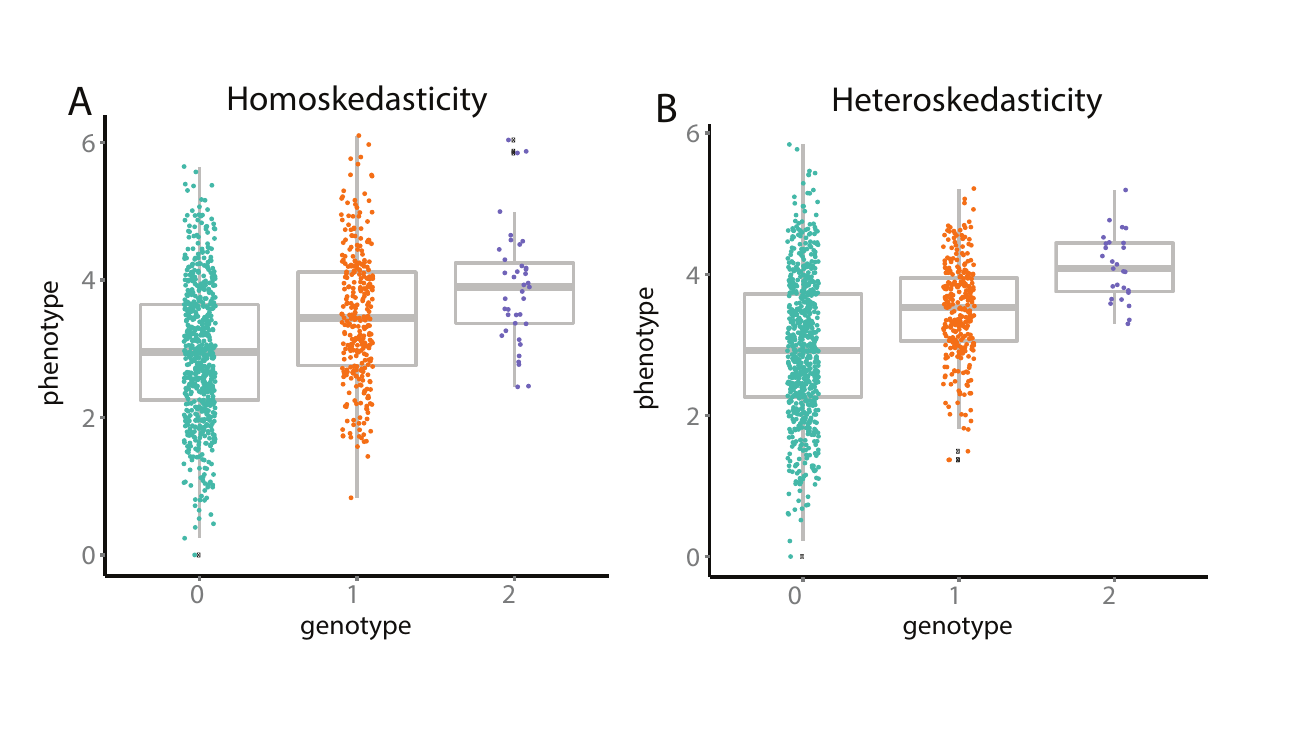}
\caption{\bf {Example of heteroskedasticity for biallelic variation.} \textmd{The x-axis is genotypes represented as the number of copies of the minor allele. 
The y-axis is the quantitative trait values across individuals sampled from a population. Panel A: \emph{Homoskedasticity}, where the variance of individuals with each of the three genotypes are equal. Panel B: \emph{Heteroskedasticity}, where the variance of individuals with each of the three genotypes are different. The data were generated from a simulation population of size $1000$, with minor allele frequency, $\pi_{maf} = 0.2$, and each genotype group was plotted with jitter along the x-axis to show data density within each genotype.}
}
\center 
\label{fig:het_example}
\end{figure}

Methodologically, detecting vQTLs is performed using statistical tests for 
\emph{heteroskedasticity}. Heteroskedasticity refers to the circumstance in which the variance of a response variable---here, a quantitative trait---is unequal across the range of values of a covariate such as genotype (Fig.~\ref{fig:het_example}). In the case of vQTLs, the quantitative traits can be gene expression levels, methylation levels, or hip-to-waist ratio; covariates may be defined by discrete or continuous variables such as genotype, sex, age, or ancestral population. Here, we develop and validate a robust statistical test for discovering variance effects in genomic analyses.

Three methods widely used in the genetics literature to identify vQTLs
are the Levene and Brown-Forysthe tests \cite{schultz1985levene,
  brown1974small}, and the correlation least squares (CLS)
test~\cite{Brown2014}.  While these tests are standard in the
heteroskedastic literature, they each have drawbacks when applied to
genomic data. The Levene and Brown-Forsythe tests both require
categorical covariates, not allowing continuous covariates such as
imputed SNPs, age, or methylation levels. These methods sacrifice
statistical power by avoiding assumptions about the functional form of
the heteroskedastic effects, allowing the variance across the
covariate-defined groups to change in a non-monotone way. Furthermore,
because they lack a parametric model, it is not possible to jointly
model possibly confounding effects. The CLS addresses all of these
drawbacks, allowing continuous covariates and additional possibly
confounding covariates such as population structure. CLS makes the
assumption that the variance of the heterozygotes is intermediate to
the variance of the two homozygotes, which improves statistical
power. However, because CLS is computed using two sequential point
estimates of parameters---neither of which incorporate
uncertainty---CLS is prone to overfitting.

In this work, we develop a Bayesian test for heteroskedasticity (BTH),
a statistical framework to enable the genome wide identification of
vQTLs or, more generally, variance effects. The model is based on
Bayesian heteroskedastic linear regression, and allows continuous or
discrete covariates and mean effects, and allows the joint modeling of
confounding covariates. We explicitly account for uncertainty in mean
effects, variance effects, and scale. We develop a fast and robust
method to approximate the test statistic using Laplace
approximations. To validate our approach, we perform simulations
across a wide range of hypothetical scenarios in association studies,
and compare the results of variants of BTH against the
state-of-the-art methods. Then we apply BTH to a number of real
genomic studies to show the impact on functional genomic analyses and
characterize the role of BTH in identifying variance QTLs relative to
related methods.

\section*{Results}

\subsection*{Bayesian Test for Heteroskedasticity (BTH)}

We developed a test for variance QTLs based on a Bayesian heteroskedastic linear regression model. In particular, we model a continuous trait $y \in \Re^n$ with a Gaussian distribution, where both the mean and variance parameters are a function of the covariate $x \in \Re^n$,
\begin{eqnarray}
y_i &\sim& {\mathcal N}(\beta_0 + \beta x_i, \sigma^2 \alpha^{ - x_i}),\\
\beta_0 &\propto& 1, \\
\beta &\sim& {\mathcal N}(0, \nu^{-1}),\\
\sigma^2 &\sim& InvGa(\theta_1, \theta_2) \\
\log(\alpha) &\sim& Cauchy(0, \nu).
\end{eqnarray}
Here, $\beta_0$ is the regression intercept, $\beta$ is the regression coefficient (or the \emph{mean effect size}), $\sigma^2$ is the residual variance, and $\alpha$ is the heteroskedastic effect. When $\alpha = 1$, the variance of the response is not a function of the covariate, whereas when $\alpha \neq 1$, the variance term is covariate-dependent. We put priors on each of these parameters in order to incorporate biologically appropriate and computationally tractable forms of uncertainty in the test; see Supplement for details.

Using this model, we computed Bayes factors~\cite{kass1995bayes} (BF) to compare the likelihood of the data under the null hypothesis ($H_0$, $\alpha = 1$) with the likelihood of the data under the alternative hypothesis ($H_A$, $\alpha \neq 1$). In particular, for each application of the model (e.g., one covariate $\bx$ and one quantitative trait $\by$ across $n$ individuals), the Bayes factor has the form
\begin{equation}
BF(\by,\bx) = \frac{\Pr(\by | \bx, (H_A, \alpha \neq 1)}{\Pr(\by | \bx, (H_0, \alpha = 1))}.
\end{equation}
We compute this BF by marginalizing over the mean effect size $\beta$ in closed form and then approximating the resulting multivariate integral using a multivariate Laplace approximation similar to the integrated nested Laplace approximation (INLA) method~\cite{rue2009approximate,ruiz2012direct} (Supplementary Materials). The BFs provide a measure of the heteroskedasticity of the association
between a covariate and a phenotype of interest under certain
assumptions, which we examine carefully in the simulations.

To quantify the global false discovery rate of the BFs, we designed
and performed permutations of the covariate - trait pair such that any
mean effects are maintained but variance effects are removed
(Supplementary Materials). Furthermore, we generate a distribution of
$BF_{perm}$s corresponding to an instance of the data set in which the
variance of the phenotype is independent of the covariate. Thus, we
are able to compute FDRs by considering, for any BF threshold $t$,
\begin{equation}
FDR(t) = \frac{|\{BF_{perm} | BF_{perm} > t\}|} {|\{BF | BF > t\}|},
\end{equation}
which captures the ratio of the number false positives versus the
number of false positives and true positives for threshold $t$ across
all tests.  We use the FDR-calibrated BF thresholds to discover
heteroskedastic associations in our data, and we compare our
discoveries to the discoveries from existing tests for
heteroskedasticity at an FDR of $0.05$.

\subsection*{Available tests of heteroskedasticity.}

We compared results from our BTH against results from three tests for heteroskedasticity used in genomics applications: i) the Brown-Forsythe test~\cite{brown1974small}; ii) the Levene test~\cite{schultz1985levene,shen2012inheritance, struchalin2012VariABEL}; and iii) the correlation least squares (CLS) test~\cite{Brown2014}. 

The Levene and Brown-Forsythe tests for heteroskasticity across $k$
groups come from a similar family of ANOVA-based statistics where the
within-group variance is compared to the across-group variance. The
null hypothesis for these tests is that all groups have the same
variance. The two tests differ in that the Levene test uses mean
statistics to compute variance whereas Brown-Forsythe uses median
statistics to compute variance. The Bartlett test
~\cite{bartlett1937properties} has also been used in genomic
context~\cite{yang2012fto}.  The Bartlett test computes \emph{pooled
  variances}, or a weighted average of variance within each covariate
group assuming the marginal distribution within groups is
Gaussian. The Bartlett test accounts for possibly imbalanced group
sizes, which is relevant for low minor allele frequency (MAF)
SNPs. While similar to the Levene test, the Bartlett test is more
sensitive to departures from normality (Supplementary Methods) and is
not discussed further here~\cite{Pare2010}.

The correlation least squares (CLS) test first fits a linear
regression model to the trait and the covariate, then tests for a
correlation between the covariate and the squared residual errors of
the fitted model using Spearman rank correlation~\cite{Brown2014}. The
returned test statistics is the corresponding Spearman rank
correlation coefficient. A related class of tests use generalized
linear models in a likelihood ratio test, and include the double
generalized linear model (\textit{dglm})~\cite{dunn2012dglm,
  ronnegaard2011detecting} and the family-based likelihood ratio test
for variance (\textit{famLRTV})~\cite{cao2015family}. In recent work,
\textit{dglm} was shown to be more prone to Type 1 errors for small
deviations from normality, as with the Bartlett
test~\cite{cao2014versatile}. Similarly, \textit{famLRTV} was found to
perform similarly to the Levene test~\cite{cao2014versatile}. Thus we
did not include these two tests in further analysis.

The Levene test, which has been used in a number of biological
studies~\cite{Ayroles2015, Pare2010,yang2012fto}, makes assumptions
about the data: i) the noise is symmetric; ii) the groups are
balanced; iii) the covariate is a categorical variable; and iv) the
categories are unordered, so arbitrary functions are tested.  By using
median statistics instead of mean statistics, the Brown-Forsythe test
overcomes the assumption of symmetric noise~\cite{brown1974small}. In
genomic studies, it is often the case that functional variants will
occur at low minor allele frequencies~\cite{nelson2012abundance},
where the balanced groups assumption will be violated. On the other
hand, the CLS test assumes i) continuous or ordered covariates; ii)
linear relationship (i.e., dosage effects of the covariates); iii)
sufficient minor allele frequency (MAF). When MAF is low, the maximum
likelihood estimates from CLS will have a large standard error.

Our model for BTH makes the following assumptions: i) the noise has a
Gaussian distribution; ii) the covariate is a continous or ordered
value; and iii) the functional form of the heterskedasticity is dosage
dependent, with monotone effects on the variance. We choose to make
these assumptions to gain statistical power in identifying
heteroskedastic effects in typical genomic study data. We note that,
as with linear regression models to test for mean effects that make an
additive assumption, it is trivial to test for dominance, recessive,
or overdominance variance effects by recoding the SNP
genotypes~\cite{plink}. In contrast to the above methods, our test
incorporates uncertainty in the model parameters instead of relying on
non-robust point estimates of those parameters, integrating over all
possible mean and variance effects in both the null and the
alternative hypothesis.

We show the value of BTH with respect to these related approaches in
extensive simulations and in three real data applications. In the
simulations, for data that violate the model assumptions, we provide
prescriptive tests and transformations to enable a well-powered
application of BTH. We then apply BTH to methylation QTLs, gene
expression QTLs, and gene expression data versus non-genetic
covariates to illustrate the promise of BTH for identifying variance
effects in diverse genomic data types.

\subsection*{Simulating quantitative trait data}

To compare results from BTH with state-of-the-art tests for variance
QTLs, we generated simulated data across a range of possible scenarios
in genomic studies. We account for both discrete and continuous
covariates, many different simulation parameter settings, and
non-Gaussian distributions of the quantitative trait.

For discrete covariates, each simulated biallelic, diploid SNP $x_i\in
\{0,1,2\}$ from individual $i=\{1,... n\}$ is sampled as two
independent draws from a Bernoulli distribution with bias equal to the
minor allele frequency ($\pi_{maf}$): $x_{i} \sim Bin(2,\pi_{maf})$.
For imputed covariates, for each individual $i=\{1,... n\}$, discrete
values $z_i\in \{0,1,2\}$ are sampled from a Bernoulli distribution:
$z_i \sim Bin(2,\pi_{maf})$. Continuous data resembling imputed
genotypes are then obtained from a modified mixture of normal
distributions: $x_i = 1_{m_i = 0} \cdot |c_0| + 1_{m_i = 1} \cdot c_1
+ 1_{m_i = 2} \cdot (2 - |c_2|)$, where $c_0, c_2 \sim
\mathcal{N}(0,0.5)$ and $c_1 \sim \mathcal{N}(1,0.5)$, and $1_{\cdot}$
is the indicator function. This process ensures that the simulated
imputed genotypes are bounded by $0$ and $2$, and they represent the
expected value of the genotype, instead of the most likely
genotype~\cite{impute2}.

Given intercept $\beta_0$, effect size $\beta$, and variance
parameters $\sigma^2, \alpha^{-x_i}$, we simulated the quantitative
trait $y_i$ for individual $i$ from a Gaussian distribution, using a
linear model:
\begin{equation}\label{eq:ideal}
y_i  \sim {\mathcal N}(\beta_0 + \beta x_i, \sigma^2 \alpha^{-x_i}).
\end{equation}
This is an ideal situation, with the heteroskedastic functional form
matching that of the model used in our test.

Across simulations, we sampled covariates and quantitative traits
across various parameter settings (defaults in bold): $n =
\{\mathbf{300}, 500, 1000\}$ samples, minor allele frequencies
$\pi_{maf} = \{0.05, \mathbf{0.2}, 0.3\}$, mean effect size $\beta =
\{0, 0.2, \mathbf{0.5}, 1\}$, and the level of heteroskedasticity
$\log \alpha = \{-0.2, -0.1, 0, \mathbf{0.1}, 0.2 \}$, intercept
$\beta_0 = \{\mathbf{0},1\}$, and a fixed variance parameter $\sigma^2
= \mathbf{1.0}$.  These simulations correspond well to current eQTL
studies in sample
size~\cite{ardlie2015genotype,battle2014characterizing}, minor allele
frequencies~\cite{nelson2012abundance}, and mean effect
sizes~\cite{savolainen2013ecological,ardlie2015genotype}.

For each parameter configuration, we generated $1,000$ simulated data
sets of covariate $\bx$ and corresponding trait $\by$. For each
simulation with heteroskedasticity, we performed a single permutation
that preserves the mean effects as for FDR evaluation. We included for
comparison this permuted null simulation with identical MAF and trait
distribution (see Online Methods). Thus, each simulation result
contains $2,000$ tests, half of which are from a null distribution
that we have constructed using permutations, where the other half are
simulated to have variance effects.

\subsection*{Simulation results: ideal model, discrete covariates.} 
For discrete genotypes, we compared results from BTH against results
from the Brown-Forsythe test~\cite{brown1974small}, the Levene
test~\cite{levene1961robust}, and the correlation least square test
(CLS)~\cite{Brown2014}. We compared performance using precision-recall
curves, which quantify the proportion of true associations discovered
(x-axis: recall or statistical power) versus the proportion of
discoveries that are truly associated (y-axis: precision, or
$1-$FDR). When the curves are close to precision $=0.5$ across most
values of recall, this means that the method cannot differentiate
between non-associations and true associations in this scenario with
equal numbers of true and null associations. The closer the curves are
to precision $=1$ across values of recall, the greater the area under
the curve (AUC) is (with a maximum of one), and the better the
performance of that method.

\begin{figure}[h!]
\centering
\includegraphics[width = 0.65\textwidth]{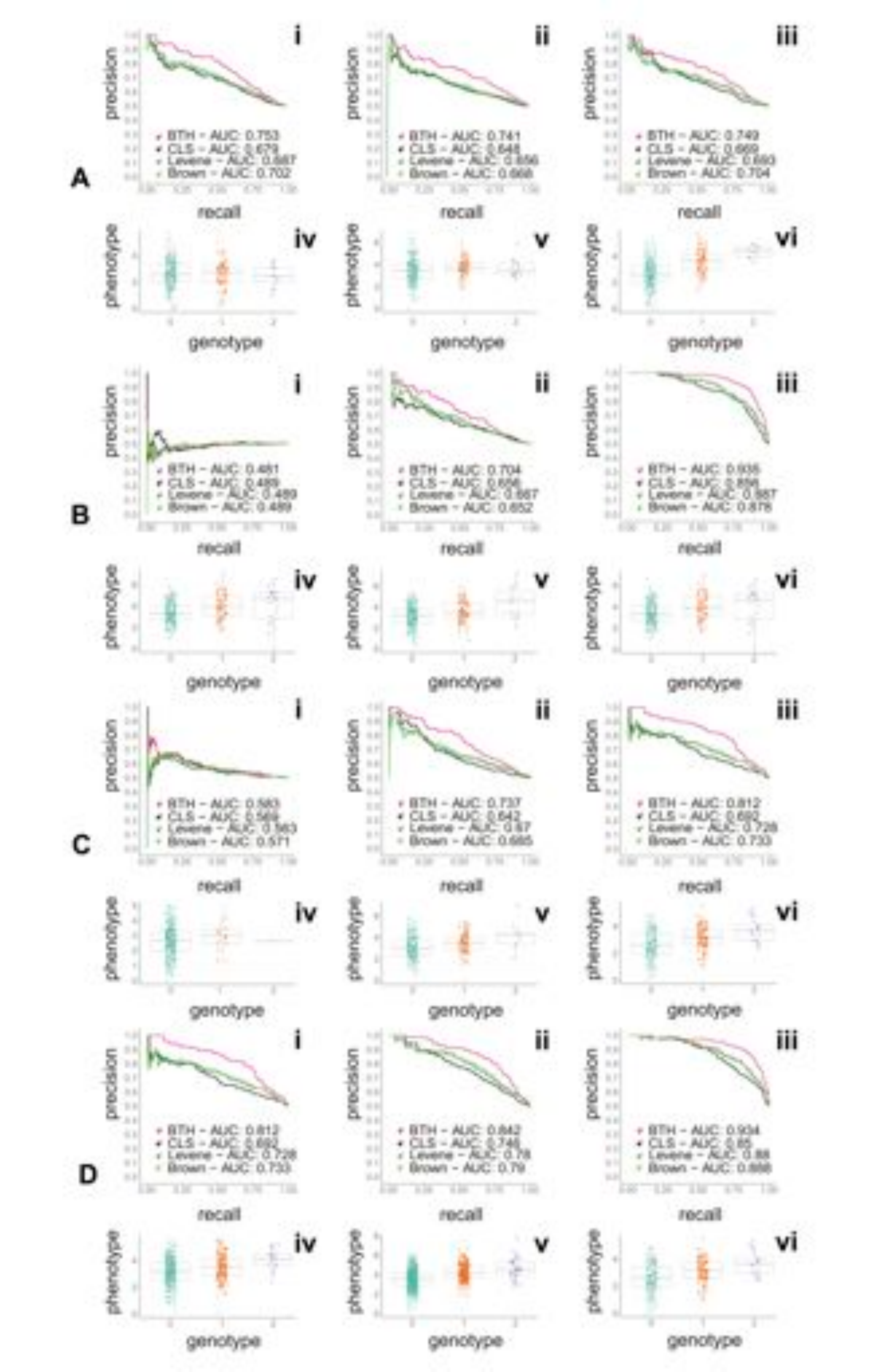}
\caption{\footnotesize {\bf Precision-recall curves comparing performance of BTH versus three other methods and example plots of underlying discrete simulated data.} 
Panel A: increasing mean effect size: $\pi_{maf} = 0.2$, $n = 300$, $\beta \in \{0,0.2,1\}$, $\log(\alpha) = 0.1$;
Panel B: increasing the variance effects:
$\pi_{maf} = 0.2$, $n = 300$, $\beta = 0.5$, $\log(\alpha) \in \{0,-0.1,-0.2\}$;  Panel C: increasing minor allele frequency: $\pi_{maf} \in \{0.05, 0.2, 0.3 \}$, $n = 300$, $\beta = 0.5$, $\log(\alpha) = 0.1$; Panel D: increasing sample size: $\pi_{maf} = 0.2$, $n \in \{300, 500, 1000\}$, $\beta = 0.5$, $\log(\alpha) = 0.1$.}
\label{fig:sim-ideal}
\end{figure}

Across possible mean effect sizes and a constant variance effect, BTH
shows consistently higher AUC compared to the other methods
(Fig.~\ref{fig:sim-ideal}Ai-Aiii). Looking across increasing mean
effects, we found that a Gaussian prior on the mean effect is robust
as mean effects increase
(Fig.~\ref{fig:sim-ideal}Ai-Aiii)~\cite{OHagan1979}.  As the variance
effects in the simulated data grow, it becomes easier for the tests to
identify these effects (Fig.~\ref{fig:sim-ideal}Bi-Biii); moreover,
the permutation appears to generate a true null
(Fig.~\ref{fig:sim-ideal}Bi) under these ideal simulation assumptions.
Here the benefit of the BTH is illustrated: when variance effect
$\log(\alpha) = -0.2$, we see at high levels of recall as much as a
$10\%$ improvement in precision (Fig.~\ref{fig:sim-ideal}Biii).

Low minor allele frequency and small sample sizes affect the AUC of
all three methods similarly (Fig.~\ref{fig:sim-ideal}Ci,Di). As MAF
and sample size increase, the AUC improves and BTH has a greater AUC
relative to the other methods (Fig.~\ref{fig:sim-ideal}Ci-Ciii,
Di-Diii). Low minor allele frequency creates imbalanced group sizes, leading to either poor or zero estimates of variance (in the case of a few or one sample in a genotype group) for a genotype. For this reason, it is important to filter out low MAF SNPs and ensure there are sufficient numbers of genotypes in each group to allow for a variance estimate when testing for heteroskedasticity with discrete or categorical covariates. 

The simulations with growing sample size highlight the benefit of Levene and Brown-Forsythe versus CLS. We note that CLS, across these ideal simulations,
appears to have generally worse performance than Levene,
Brown-Forsythe, and BTH. In particular, the AUC of BTH is significantly
greater than either Levene, Brown-Forsythe, or CLS (t-test $p\leq 1.4\times
10^{-203}$, $5.2\times 10^{-177}$, $1.6\times 10^{-287}$ respectively
in Di) with an average precision $5\%$ higher. Similarly,
for greater magnitude mean effects (Fig.~\ref{fig:sim-ideal})Ai-Aiii) and growing sample
sizes (Fig.~\ref{fig:sim-ideal}Di-Diii), the AUC of CLS is up to $0.036$ smaller than the
AUC of either Levene or Brown-Forsythe, and as much as $0.12$ smaller than
the AUC of BTH.

\subsection*{Simulation results: ideal model, continuous covariates.}

When the covariate is a continuous value---such as age, BMI, or the expected number of minor alleles for imputed genotypes---the Levene and Brown-Forsythe tests are no longer appropriate, as they assume categorical covariates. In this case, we compared our method with the CLS method, which allows a general covariate in the original linear regression and subsequent correlation test. We also applied Brown-Forsythe and Levene tests to the simulated data by rounding the continuous covariates to their nearest integer value. For imputed genotypes, this rounding process corresponds to setting the value of the covariate SNP to the most likely number of copies of the minor allele. This is an idealized imputation scenario (results on imputed genotypes below are much less straightforward), and we would not recommend this approach with real data~\cite{marchini2010genotype}.

\begin{figure}
\centering
\includegraphics[height = 0.8\textheight]{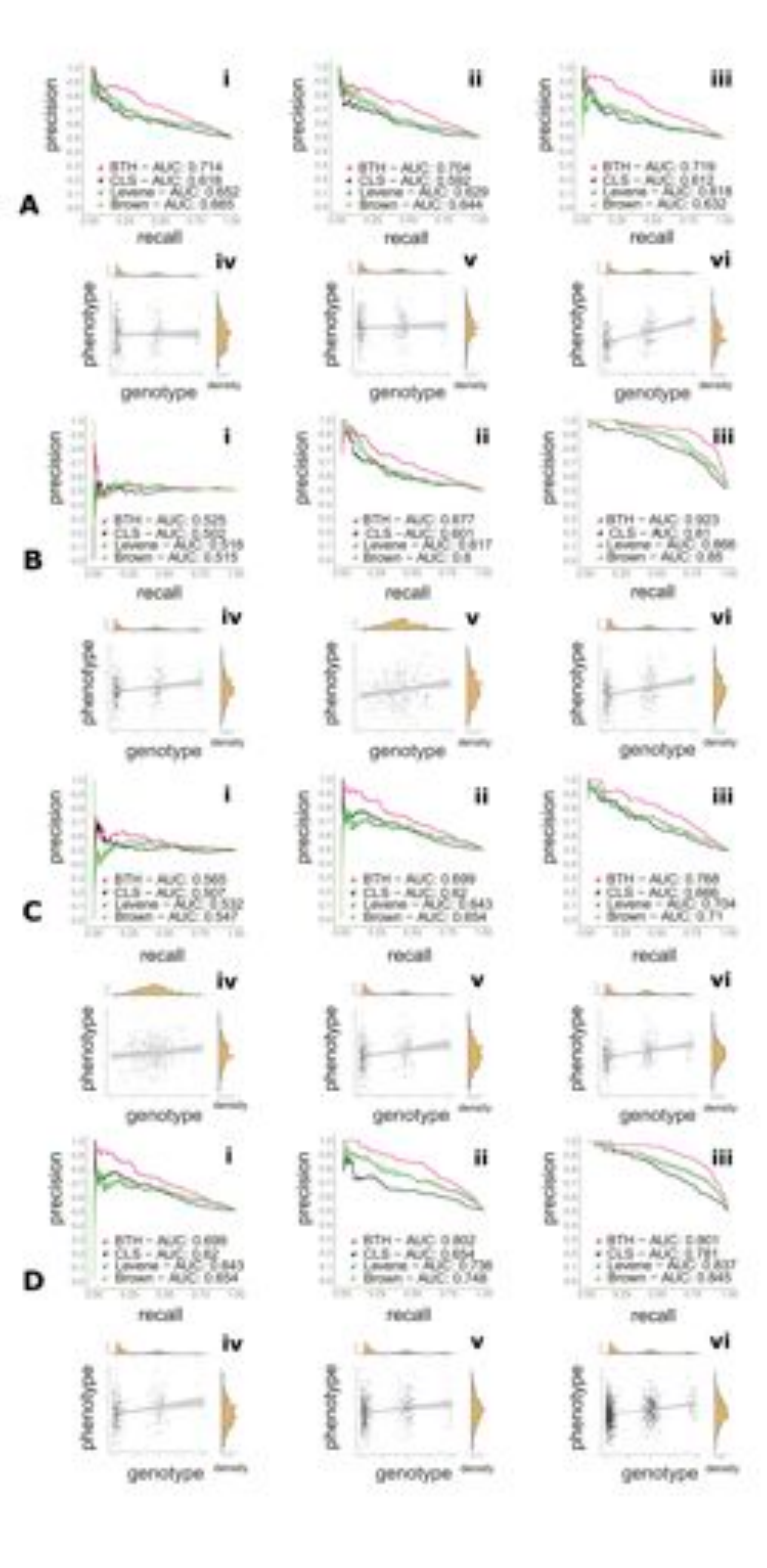}
\caption{\footnotesize {\bf Precision-recall curves comparing performance of BTH versus CLS, Levene, and Brown-Forsythe and corresponding scatter plots of one example of this scenario.} Panel A: increasing the mean effects: $\pi_{maf} = 0.2$, $n = 300$, $\beta \in \{0,0.2,1\}$, $\log(\alpha) = 0.1 $; Panel B: increasing the variance effects: $\pi_{maf} = 0.2$, $n = 300$, $\beta = 0.5$, $\log(\alpha) \in \{0,-0.1,-0.2\}$; Panel C: increasing the minor allele frequency: $\pi_{maf} \in \{0.05, 0.2, 0.3 \}$, $n = 300$, $\beta = 0.5$, $\log(\alpha) = 0.1$; Panel D: increasing the sample size: $\pi_{maf} = 0.2$, $n \in {300, 500, 1000}$, $\beta = 0.5$, $\log(\alpha) = 0.1$.}
\label{fig:sim-continuous}
\end{figure}

The results on the continuous covariate simulations echo the discrete simulation results. In particular, BTH shows a uniform improvement in AUC across all of the simulations, considering increasing mean effects (Fig.~\ref{fig:sim-continuous}A), increasing variance effects (Fig.~\ref{fig:sim-continuous}B), increasing minor allele frequency (Fig.~\ref{fig:sim-continuous}C), and increasing sample size (Fig.~\ref{fig:sim-continuous}D). We note that, despite rounding the (idealized) imputed genotypes, Levene and Brown-Forsythe continue to perform better than CLS across the simulations.

\subsection*{Simulation results: non-ideal model, discrete covariates.}

Next, we explored quantitative traits simulated from four non-ideal heteroskedastic models with discrete covariates that are motivated by residual distributions often found in genomic analyses.
\begin{enumerate}
\item \emph{Additive variance term:} a Gaussian distribution with an additive form of heteroskedasticity:

\begin{equation}\label{eq:linvar}
y_i  \sim {\mathcal N} (\beta_0 + \beta x_i, \sigma^2 + \alpha \cdot x_i).
\end{equation}
We generated data with additive variance effects to ensure that our test is able to identify different functional forms of heteroskedasticity. Note that, in this scenario, the null hypothesis corresponds to $\alpha = 0$; parameters in this category of simulation reflect this different null.

\item \emph{Log Gaussian:} the log of the trait follows a Gaussian distribution:
\begin{equation} \label{eq:log}
y_i  \sim \exp \left\{{\mathcal N}\left( \beta_0 + \beta x_i , \sigma^2 \alpha^{-x_i}\right )\right\}.
\end{equation}
Microarray data (within gene) are believed to have a log Gaussian distribution, which motivates log transformations to those data~\cite{Irizarry2003}. Untransformed log normal data, however, will naturally appear heteroskedastic because of the correlation of mean and variance in the log Gaussian distribution.

\item \emph{Gamma distributed data:} traits are generated from a gamma generalized linear model:
\begin{eqnarray}\label{eq:Gamma}
\mu_i  =  \frac{1}{\beta_0 + \beta x_i}\nonumber \\
y_i \sim Gamma\left(\mu_i, 1\right); 
\end{eqnarray}
While the exponential distribution is the continuous form of the Poisson distribution, the gamma distribution may be considered the continuous form of the negative binomial distribution, which is a discrete distribution with an additional variance parameter above the discrete Poisson distribution. Hence, we generate continuous data from the gamma distribution to simulate the continuous trait form of overdispersed Poisson counts, as might be found in mapped RNA-sequencing
data~\cite{pickrell2010understanding, marioni2008rna}.

\item \emph{Mixture of Gaussians:} traits are generated from a mixture of two Gaussian components, one heteroskedastic and one homoskedastic, with mixture parameter $\lambda = 0.4$:
\begin{eqnarray}\label{eq:MixGau}
y_i \sim \lambda {\mathcal N}(10,1) + (1- \lambda) {\mathcal N}(\beta_0 + \beta x_i, \sigma^2 \alpha^ {-x_i}).
\end{eqnarray}
We expect bimodal Gaussian traits when, for example, there is a GxG or GxE interaction. For example, if there is a mean effect for female samples at a SNP, but no corresponding mean effect for male samples, the quantitative trait will appear bimodal within genotype.
\end{enumerate}

We quantified the relative performance of the tests using precision-recall curves as above; however, caution must be used here in interpreting the relative AUC.  We may consider four possibilities (Table~\ref{tab:NullStrong}) for the simulated mean and variance effects with respect to the statistical test we perform here:
\begin{itemize}
\item \emph{Strong null:} the simulated mean effects $\beta = 0$ and the simulated variance effects $\log(\alpha) = 0$;
\item \emph{Weak null:} the simulated mean effects $\beta \neq 0$ and the simulated variance effects $\log(\alpha) = 0$;
\item \emph{Weak alternative:} the simulated mean effects $\beta= 0$ and the simulated variance effects $\log(\alpha) \neq 0$;
\item \emph{Strong alternative:} the simulated mean effects $\beta \neq 0$ and the simulated variance effects $\log(\alpha) \neq 0$.
\end{itemize}

\begin{table}
\centering
\begin{tabular}{l | c | c }
Hypothesis & strong & weak \\
\hline \hline
Null  & $\beta = 0$ and $\log(\alpha) = 0$ & $\beta \neq 0$ and $\log(\alpha) = 0$\\ 
Alternative & $\beta \neq 0$ and $\log(\alpha) \neq 0$ & $\beta = 0$ and $\log(\alpha) \neq 0$ \\
\end{tabular}
\caption{{\bf Different hypotheses tested in various data scenarios.} The BTH model integrates over the mean effect size, $\beta$, testing the union of the weak and strong alternative hypotheses against the union of the weak and strong null hypotheses.}
\label{tab:NullStrong}
\end{table}

These definitions become important when discussing the log Gaussian
and gamma simulations: for both distributions, the variance is a
function of the mean, inducing an explicit relationship between the
two. In other words, when there are mean effects, $\beta \neq 0$, this
will present as variance effects in these tests.  The BTH model
integrates over the mean effect size, $\beta$, testing the union of
the weak and strong alternative hypotheses against the union of the
weak and strong null hypotheses. Moreover, in the permutations, we
specifically remove variance effects while maintaining mean
effects. These design decisions lead to different behavior of the test
on these simulations from non-ideal genomic scenarios.

For the non-ideal simulations, we simulated data both from the strong
alternative ($\beta \neq 0$, $\log(\alpha) \neq 0$;
Fig.~\ref{fig:nonideal}, first column) and the weak null ($\beta \neq
0$, $\log(\alpha) = 0$; Fig.~\ref{fig:nonideal}, second column). The
weak null simulation ideally will look like the strong null
simulations; however, for the log Gaussian and gamma simulations, all
tests differentiate the weak null and the strong null as an artifact
of the data distribution linking mean and variance effects. This
phenomenon may be seen in the results by comparing the AUC of the
strong alternative simulations with the weak alternative simulations:
for the gamma simulations, the four tests have nearly identical AUCs
regardless of the true value of the variance effects $\alpha$. This
suggests that the performance in the strong alternative simulations is
due to mean effects. We verified this by considering simulations from
the weak alternative (i.e., $\log(\alpha) \neq 0$, $\beta = 0$),
finding that all of the tests fail to detect signs of
heteroskedasticity in the gamma simulations (Fig. S19Aii, S19Bii,
S19Cii). Similarly, in the untransformed log Gaussian simulations,
test performance on the weak alternative scenario is close to that for
the strong null (Fig. S16Aii,iv).

For the additive variance effects simulations and the bimodal
distributed simulations, we find that the weak null simulations are
appropriately unable to differentiate the weak null from the true null
simulations (Fig.~\ref{fig:nonideal}Ai-Aiv; Di-Div). Moreover, for the
bimodal distributed simulations, BTH had the most substantial gains in
AUC relative to the other three methods, all of which had noticeably
worse performance than in the ideal unimodal simulations. We further
study departures from the ideal distributions below in the real data
applications.

\subsection*{Simulation results: classifying and transforming non-ideal data distributions.}

\begin{figure}
\centering
\includegraphics[width = 0.6\textwidth]{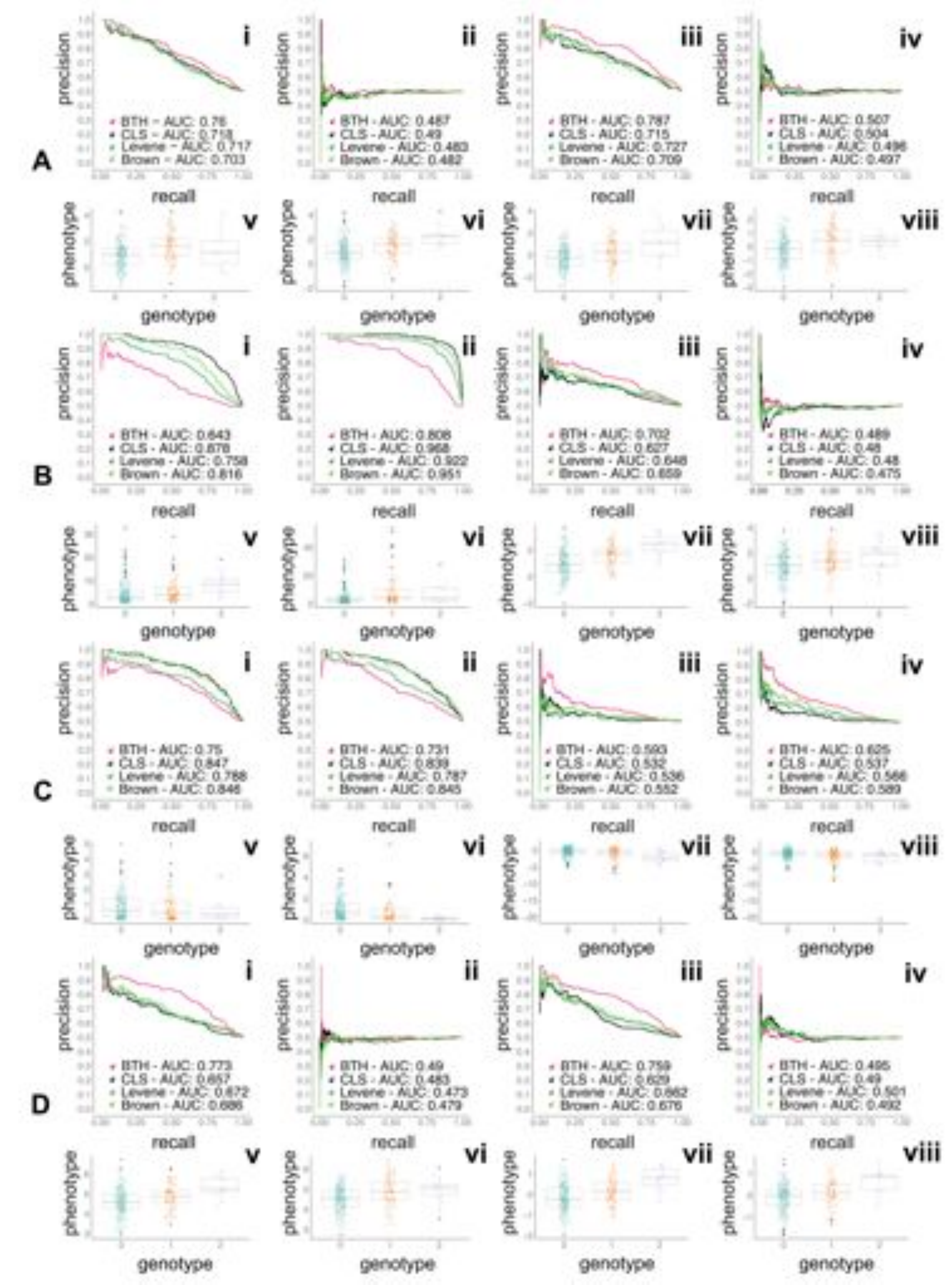}

\caption{\footnotesize {\bf Precision-recall curves illustrating departures from the idealized model for both the strong alternative and the weak null models, and their respective transformations.} Across all panels $\pi_{maf} = 0.2$, $n = 300$, $\beta_0 = 1$, $\beta = 0.5$, $\sigma^2 = 1$. Panel A: Additive variance model:  
i) No transformations, $\alpha \neq 1$, $\beta \neq 0$ (strong alternative); ii) No transformations, $\alpha = 0$, $\beta \neq 0$ (weak null); iii) Mean centered, $\alpha = 1$, $\beta \neq 0$ (strong alternative); iv) Mean centered,  $\log(\alpha) = 0$, $\beta \neq 0$ (weak null);
Panel B: Log Gaussian model: i) No transformations, $\log(\alpha) \neq 0$, $\beta \neq 0$ (strong alternative); ii) No transformations, $\log(\alpha) = 0$, $\beta \neq 0$ (weak null); iii) Log transformation, $\log(\alpha) \neq 0$, $\beta \neq 0$ (strong alternative); iv) Log transformation, $\log(\alpha) = 0$, $\beta \neq 0$ (weak null);
 Panel C: Gamma model:  
i) No transformations, $\log(\alpha) \neq 0$, $\beta \neq 0$ (strong alternative); ii) No transformations, $\log(\alpha) = 0$, $\beta \neq 0$ (weak null); iii) Mean centered square root transform, $\log(\alpha) \neq 0$, $\beta \neq 0$ (strong alternative); iv) Mean centered square root transform,  $\log(\alpha) = 0$, $\beta \neq 0$ (weak null);
 Panel D: Bimodal model:
i) No transformations, $\log(\alpha) \neq 0$, $\beta \neq 0$ (strong alternative); ii) No transformations, $\log(\alpha) = 0$, $\beta \neq 0$ (weak null); iii) Mean centered, $\log(\alpha) \neq 0$, $\beta \neq 0$ (strong alternative); iv) Mean centered,  $\log(\alpha) = 0$, $\beta \neq 0$ (weak null). }
\label{fig:nonideal}
\end{figure}

To address the problem of distributional misspecification of the model, we developed a statistical classifier that takes as input the $x$ and $y$ vectors (covariates and traits, respectively) and returns the probability of each of seven distributions within and across groups for discrete covariates and across values for continuous covariates (see Supplemental Methods, Fig. S34, S35 and Tables $5-8$). Given a distribution classification for a particular covariate-trait test, we then suggest a specific data transformation to encourage a $0.5$ recall for the weak null simulations (i.e., mean effects but no explicit variance effects). In particular, when the data appear to have a log Gaussian distribution, we suggest a log transformation  (Figure~\ref{fig:nonideal}B); when the data appear Gamma distributed, we suggest a mean-centered square root transformation (Figure~\ref{fig:nonideal}C).

We compared the transformed strong alternative simulations ($\beta
\neq 0$, $\log(\alpha) \neq 0$; Fig.~\ref{fig:nonideal}, third
column), and found that BTH uniformly had the largest AUC across the
four methods. We also compared results on the transformed weak null
simulations ($\beta \neq 0$, $\log(\alpha) = 0$;
Fig.~\ref{fig:nonideal}, fourth column). These show that the
transformation eliminates the mean effect discoveries in all but the
gamma simulations (Fig.~\ref{fig:nonideal}C); in the gamma
simulations, variance effects are nearly removed across the four
methods. We explore gamma-distributed data further in the methylation
data analysis below.

\
\

\subsection*{1000 Genomes Project Methylation Study Data}

We applied BTH and the alternative tests for variance effects to a genome scale differential DNA methylation study~\cite{heyn2013dna} to find variance methylation QTLs (meQTLs). These data consist of DNA methylation levels at $485,577$ CpG sites across the human genome using the Infinium HumanMethylation450 BeadChip platform (Illumina) in lymphoblastoid cell lines (LCLs) from 288 individuals -- 96 American with Northern and Western European ancestry, 96 Han Chinese, and 96 Yoruban. Following previous work, we removed CpG probes of poor quality or with common mutations. We used the $\beta$ values from the methylation arrays at $406,021$ CpG sites for analysis.
Genotype information for these individuals are available
from HumanHap550k and HumanHap650k SNP arrays (Illumina) at the GEO accession numbers GSE24260~\cite{kalari2010copy} (192 individuals) and GSE24274~\cite{niu2010radiation} (96 individuals). We removed eight individuals that did not have methylation data, and combined the genotypes from $280$ individuals with $170,063$ SNPs common to both genotype platforms and without missing data. 
For each CpG site, we tested for association with cis-SNPs, defined as SNPs within 10KB of the CpG site~\cite{heyn2013dna,Bell2011}.
We evaluated the global FDR of our association results using a single permutation of the methylation data. Significance was assessed using a global FDR and FDR stratified by MAF using permutations as above~\cite{sun2006} (Supplementary Methods Table $S1$).

\begin{figure}
\centering
\includegraphics[scale = 0.20]{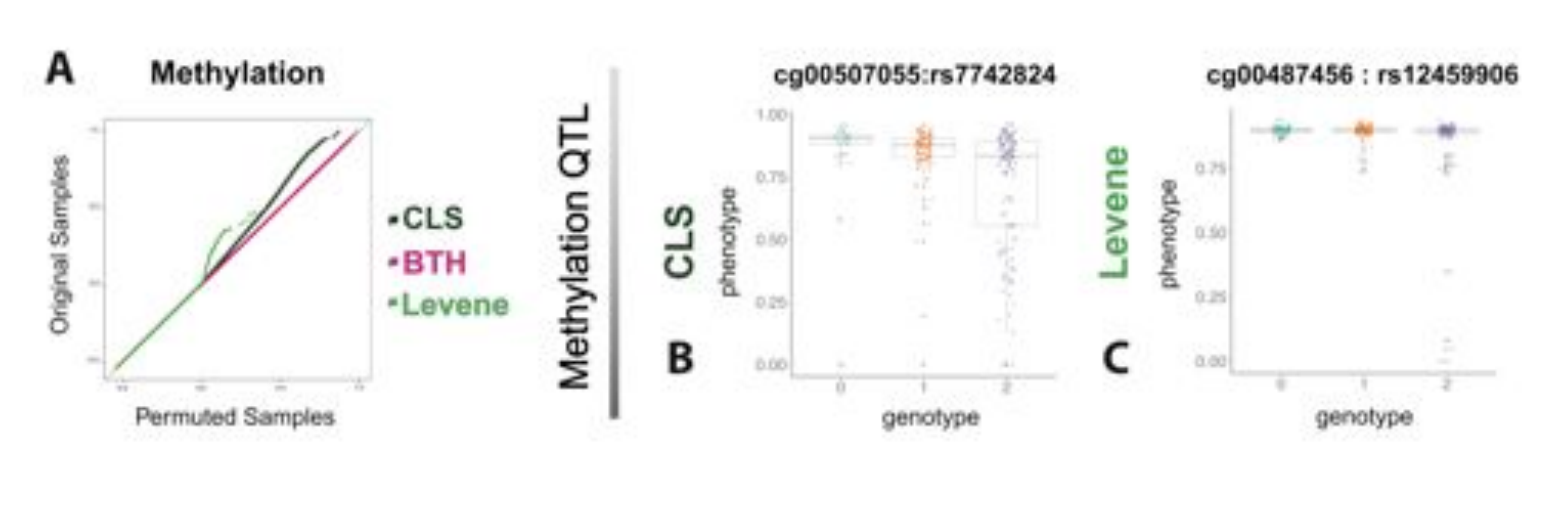}
\\
\caption{Results from tests of association between DNA methylation levels and genotype at an FDR of  $0.05$. Panel A: Quantile-quantile plot of Bayes factors in permuted (x-axis) and unpermuted (y-axis) data using the BTH, Levene and CLS methods; Panel B: False positive discovery using the CLS test; Panel C: False positive discovery using Levene. }
\label{fig:summaryDiscoveryMET}
\end{figure}

Using our distribution classifier, we found that most of the methylation level traits were gamma distributed (Supplemental Tables $S6$ and $S7$).
BTH does not find any significant variant-mediated associations between SNPs and methylation levels at CpG sites at a global FDR of $0.05$ 
and a MAF-stratified FDR of $0.05$ (Fig.~\ref{fig:summaryDiscoveryMET}A; Table $S1$).
In contrast, CLS found $391$ significant associations and the Levene test found $2123$ associations (Fig.~\ref{fig:summaryDiscoveryMET}A; global FDR $\leq 0.05$). However, in these discoveries, a large majority of the the methylation levels were bimodal or multimodal, with unimodal traits making up $8.95$\% of the discoveries from CLS and $0.38$\% of the discoveries from the Levene test (Fig.~\ref{fig:summaryDiscoveryMET}B,C). We hypothesize that the bimodal distribution of methylation values with respect to genotype is due to epistatic effects. BTH, on the other hand, is robust to bimodal deviations from the unimodal Gaussian distribution, and does not detect these candidates for epistatic effects at an FDR $\leq 0.05$. 

We tested for variance meQTLs without transforming the methylation data under the assumption that a single SNP will not have both mean and variance effects on methylation levels at a single CpG site. BTH detected no variance associations, meaning that false positives due to confounding effects were not apparent in the data. Had there been 
discoveries for BTH, we would have repeated the test with the appropriately transformed data using a square root transform (Figure $S19$). 
Tests of heteroskedasticity are sensitive to data distribution, such as skewed methylation level data~\cite{feinberg2010stochastic,feinberg2010personalized, jaffe2012significance}, but we do not quantile normalize these methylation data in order to prevent removing potential variance effects.
\subsection*{Cardiovascular and Pharmacogenetics (CAP) study.}

We applied BTH to test for variance effects between imputed genotypes and gene expression levels from the Cardiovascular and Pharmacogenetics (CAP) study. Gene expression values for $10,195$ genes in lymphoblastoid cell lines (LCLs) from $480$ Caucasian individuals were assayed on human microarray platforms~\cite{mangravite2013statin}. Genotypes were assayed using genotyping arrays and subsequently imputed using IMPUTE2~\cite{impute2,impute2_1,mangravite2013statin} to yield $33,386,856$ total markers across the 22 autosomal chromosomes. We removed SNPs with MAF below $0.05$.

The preprocessing of gene expression data for testing of variance eQTLs is somewhat different than the preprocessing for mean effect eQTLs \cite{irizarry2003RMA}. To test for variance eQTLs, we log transform the microarray gene expression data so they do not have a log normal distribution, we control for outliers, and we control for known (directly measured) and unknown (inferred) confounders. Specifically, we 1) $\log_2$ transform each gene expression level; 2) project each set of gene expression levels to the quantiles of the empirical gene distribution; and 3) control for known covariates (age, sex, batch), the first two principal components of the expression matrix, and the first two principal components of the genotype matrix (see Supplemental Methods). After empricial quantile normalization \cite{Brown2014}, each gene has exactly the same distribution across all samples, and a visual analysis of a QQ-Plot confirms the empirical distribution deviates little from a normal distribution (Figure S29).
After preprocessing the genotype and gene expression data, we performed association mapping between each gene and the cis-SNPs local to that gene; here, cis-SNPs are defined to be $\leq 200$Kb from the gene transcription start or end site~\cite{pickrell2010understanding}. There were $9,862$ genes with at least one cis-SNP in these data, and, on average, each gene had $847$ cis-SNPs. We computed the test statistic for the putative association between each cis-SNP gene pair with these processed gene expression data~\cite{pickrell2010understanding}.

These gene expression analyses find three variance QTLs using CLS and
six variance QTLs using BTH (FDR$\leq 0.05$;
Fig.~\ref{summaryDiscoveryCOV}I). This could be due to the fact that
the distributional properties of the processed CAP data set conform to
the assumptions of the statistical tests, namely normal residuals or
mild departures from normality (i.e., linear variance, exponential
mean, or log normal residuals; Table $S4$) as compared to the
generally gamma-distributed residuals in the methylation data. In
particular, BTH discovers variance effects for:
\begin{itemize}
\item RNA-binding protein \textit{C14orf166} at SNP \textit{rs1953879} (Figure~\ref{summaryDiscoveryCOV}I-A);
\item the oligosaccharyltransferase complex subunit \textit{OSTC} or \textit{DC2} at SNP \textit{rs145858690} (Figure~\ref{summaryDiscoveryCOV}I-B);
\item the type B histone acetyltransferase gene $\textit{HAT1}$ at four SNPs: \textit{rs4635498}, \textit{rs10445763}, \textit{rs75847248}, and \textit{chr2:172709927} (Figure~\ref{summaryDiscoveryCOV}I-C).
\end{itemize}

The association between gene \textit{C14orf166} and SNP
\textit{rs1953879} is particularly interesting. Gene
\textit{C14orf166} is an RNA-binding protein involved in the
modulation of mRNA transcription by Polymerase II. The role of
\textit{C14orf166} in development and progression of several human
cancers -- cervical cancer \cite{zhang2015c14orf166}, brain tumors
\cite{wang2007hupo}, and pancreatic cancer \cite{perez2014hcle}---is
well characterized. 
Moreover, the locus associated with the gene,
\textit{rs1953879}, is located within the protein coding abhydrolase
domain containing the 12B gene, \textit{ABHD12B}. The protein
associated with \textit{ABHD12B} is thought to be involved in metabolic
processes related to acylglycerol lipase activity and has recently
been associated with obesity related phenotypes~\cite{ABHD12Bobesity}.
Its paralog, \emph{ABHD12}, catalyzes the main endocannabinoid lipid
transmitter and is associated with a number of physiological processes
and complex phenotypes including mood, appetite, addiction behavior,
and inflammation \cite{ABHD12infla}. This is interesting in the
context of prior work on variance effects due to epistasis at the
BMI-regulating \emph{FTO} locus~\cite{Pare2010}.

Similarly, CLS discovers variance effects for:
\begin{itemize}
\item  the Jumonji domain-containing protein 1C \textit{JMJD1C} at SNP \textit{rs113764414} (Figure~\ref{summaryDiscoveryCOV}I-D);
\item  the chromatin modifying protein 6 \textit{CHMP6} at SNP \textit{rs7217916} (Figure~\ref{summaryDiscoveryCOV}I-E);
\item the protein coding aryl hydrocarbon receptor interacting protein-like \textit{AIPL1} gene at SNP \textit{chr17:6379486} (Figure~\ref{summaryDiscoveryCOV}I-F);
\end{itemize}
Among the CLS results, the \emph{AIPL1} association with the $chr17:6379486$ locus ranks in the top $0.02\%$ most significant Bayes factors generated by BTH. 

\subsection*{CAP study covariates.}

Next, we applied BTH to test for a heteroskedastic relationship
between gene expression levels and known covariates collected on the
individuals enrolled in the CAP study~\cite{cao2015family}. In
particular, we considered sample age, sex, BMI, and smoking
status. For non-binary covariates (age and BMI), we normalized the
values --- dividing by the maximum of each so that the covariate had a
maximum value of one --- for stability of parameter estimation; this
does not change interpretation of our results. Both sex and smoking
status are binary covariates, so the application of BTH is equivalent
to testing for differential variance across the binary
covariate. Overall, these data contain 46\% females and 87\%
non-smokers.

\begin{figure}
\centering
\includegraphics[width=0.9\textwidth]{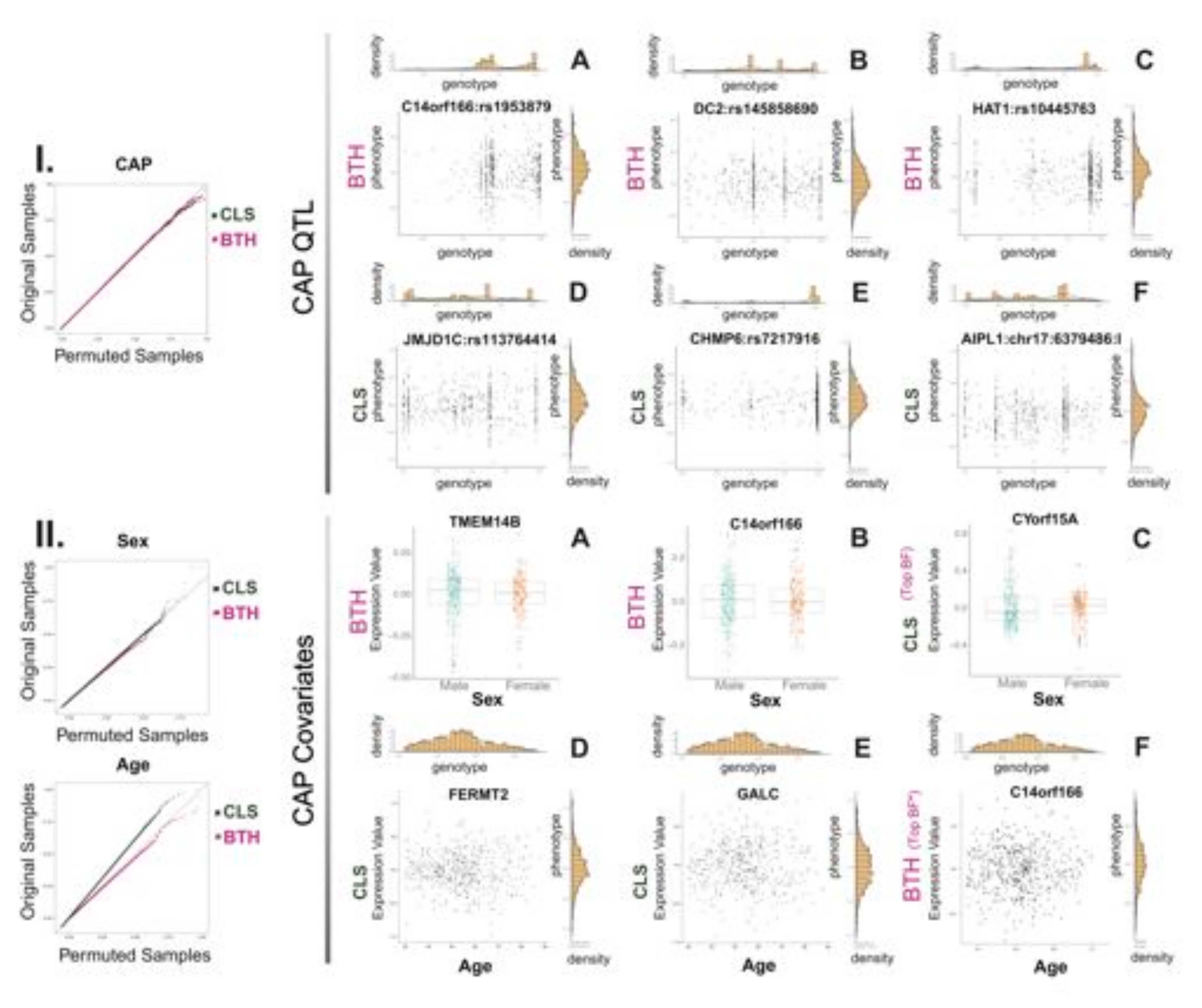}
\caption{Variance QTLs uncovered by BTH and related tests in CAP gene expression and CAP covariate datasets at FDR of $0.05$$^{*}$. Panel I -- Quantile-quantile plot of Bayes factors in permuted (x-axis) and unpermuted (y-axis) CAP gene expression data using the BTH and CLS methods: Panel A: \emph{C14orf166} versus rs195387 (BTH); Panel B: \emph{DC2} versus rs14585869(BTH); Panel C:  \emph{HAT1} versus rs10445763(BTH); Panel D: \emph{JMJD1C} versus rs113764414 (CLS); Panel E: \emph{CHMP6} versus rs7217916 (CLS); Panel F: \emph{AIPL1} versus chr17:6379486 (CLS).
Panel II -- Quantile-quantile plot of Bayes factors in permuted (x-axis) and unpermuted (y-axis) CAP covariate data using the BTH and CLS methods
covariate sex versus \emph{TMEM4B} (BTH); Panel B: covariate sex versus \emph{C14orf166} (BTH); Panel C: covariate sex versus \emph{CYorf15A} (CLS);  Panel D: covariate age versus \emph{FERMT2} (CLS); Panel E: covariate age versus \emph{GALC} (CLS); Panel F: covariate age versus \emph{C14orf166} (BTH; $*$ top unpermuted association according to BTH, occurring at an FDR higher than $1$ - $3$ permuted results rank higher).}
\label{summaryDiscoveryCOV}
\end{figure}

In these data, we found five significant associations using BTH and
CLS (FDR$\leq 0.05$; Fig.~\ref{summaryDiscoveryCOV}II; Supplementary
Table $S5$). Among the five significant associations, three
corresponded to sex specific variance control (identified using BTH),
and two corresponded to age specific variance control (identified
using CLS). In particular, BTH discovers variance effects of sex in
the transmembrane protein 14B, \textit{TMEM14B}, in the gene
\textit{C14orf166}. The association between sex and gene
\textit{CYorf15A} is the third most significant Bayes factor, although
it is not significant at an FDR$\leq 0.05$. CLS discovers variance
effects of age in the genes \textit{FERMT2} and \textit{GALC}. Among
these, \textit{FERM2} or \textit{Kindlin-2} is known to interact with
beta catenin and is associated with the integrin signaling pathway,
cell adhesion and mutagenesis \cite{KindlinEx}.

The sex-linked gene association is interesting: \textit{CYorf15A} is located on the Y chromosome and is thus not expressed in females (Fig.~\ref{summaryDiscoveryCOV}II - C). This association then is expected: the variance in females is only due to measurement error, whereas the variance in males is due in part to true expression differences. Thus, this association acts as a control for our test. However, this example shows the conservative nature of our testing pipeline, in that we did not identify all Y-linked genes in these data. The other two genes, on the other hand, are located on autosomal chromosomes.

\section*{Discussion}

We presented a Bayesian test for heteroskedasticity (BTH) that allows
for continuous covariates, and incorporates uncertainty in estimates
of mean and variance effects to robustly test for variance QTLs and
other types of variance effects. We showed the promise of our approach
on extensive simulations within and in violation of the assumptions in
our model, and describe a prescriptive procedure to ensure a well
powered application of our model to diverse genomic and epigenetic
study data. We emphasize that, although we are mainly focused on
variance effects of genotype on quantitative traits, this approach may
be used broadly in testing for heteroskedastic associations between
two sample measurements, and we illustrate this application by
discovering meaningful associations between non-genetic covariates and
gene expression data.

Our results show that BTH is more conservative and less
sensitive to multi-modal distributions than both CLS and the Levene
test, as found in the methylation data. In scenarios where the real
data is close in distributional form to the modeling assumptions, as
in the gene expression data, BTH finds similar numbers of associations
as CLS. Moreover, the largest ranked $-\log_{10}(p-values)$ from CLS
correspond to the most significant Bayes factors from BTH, meaning
that the tests generally agree in the real data scenarios. 

The lack of substantial numbers of results from BTH in the genomic
studies raise an important discussion point. In particular, the
signature of gene x gene or gene x environment epistatic interactions
may show up as a bimodal distribution of the trait: consider the
distribution of a trait that has an eQTL with mean effect in women but
not in men~\cite{Muther2011}. We note that our statistical test was
robust to deviations from unimodality, but CLS and Levene were not,
making the purpose of these tests somewhat orthogonal. Thus, to
identify candidate epistatic associations with bimodal distributions,
CLS and Levene are the appropriate methods to use and the data should
not be quantile normalized; on the other hand, to avoid bimodal
associations, we have shown that our method is preferable in terms of
statistical power. We also hypothesize that the permutations that are
used for these tests, while appropriate, lead to conservative
estimates of FDR, which impacted all of the statistical tests
calibrated using permutations.

The lack of power in the variance QTL studies was clear: as opposed to
mean effects, we found no variance effects on methylation and six
significant variance effects on gene expression levels. We propose
that these measurements of cellular traits are inappropriate
candidates for variance QTLs because variance effects are generally not
across individuals but instead across cells within an individual, as
was identified in previous studies~\cite{wills2013single}.
In particular, the effect of a variance eQTL will be to impact the variability of gene expression or methylation levels across the sample cells. The measurement of these cellular traits levels, however, are performed on tens of thousands of cells, and quantify the average expression or DNA methylation levels across those cells. Thus, in order to identify variance eQTLs and variance meQTLs, different types of data must be considered such as single cell RNA-sequencing data~\cite{wills2013single} or resampled RNA-sequencing data to estimate within-sample variance~\cite{auer2010statistical}, but neither of these domains are currently feasible to discover variance QTLs.

\section*{Methods}

\subsection*{Bayesian Test for Heteroskedasticity (BTH).}

The observed data are two vectors, $y \in \Re^{n}$ (quantitative trait) and $x \in \Re^{n}$ (covariate). For each sample $i= \{1,\ldots, n\}$, we model the quantitative trait as a function of the covariate as follows: 
\begin{equation}
y_i \sim \beta_0 + \beta x_i + {\mathcal N}(0, \sigma^2 \alpha^{-x_i}).
\end{equation}
The parameters for the BTH model use appropriate priors:
\begin{eqnarray*}
\beta \: | \: \nu \sim& {\mathcal N}(0, \nu^{-1}) &\mbox{  Gaussian prior} \\
\sigma^2 \: | \: \theta_1, \theta_2 \sim& InvGa(\theta_1, \theta_2) & \mbox{  Inverse gamma prior}\\
\beta_0 \propto& 1 & \mbox{  Uniform prior}\\ 
\log(\alpha)  \: | \: x_0, \gamma \sim& Cauchy(x_0, \gamma)  &\mbox{  Cauchy prior}\\
\end{eqnarray*}
where $x_0 = 0$, centering the Cauchy distribution at $0$, $\nu = 5$, $\theta_1 = 1$, $\theta_2 = 2$, and $\gamma = 1$.

BTH computes the likelihood of the alternative hypothesis versus the likelihood of the null hypothesis~\cite{kass1995bayes}:
\begin{itemize}
\item $H_0$, the null hypothesis where $\alpha = 1$ or, equivalently, $\log(\alpha) = 0$;
\item $H_A$, the alternative hypothesis where $\alpha \neq 1$ or, equivalently, $\log(\alpha) \neq 0$.
\end{itemize}
We computed Bayes factors $BF(y,x) = \frac{\Pr(y \big| x, H_A)}{\Pr(y \big| x, H_0)}$ as follows:\
\begin{equation}
 \frac{\int_0^{\infty}\int_0^{\infty}\int_0^{\infty} \Pr(y \big| x; \beta_0, \sigma^2, \alpha) \Pr(\beta_0,\sigma^2) \Pr(\alpha) d \beta_0 d \sigma^2 d \alpha}{\int_0^{\infty}\int_0^{\infty} \Pr(y\big| x; \beta_0, \sigma^2, \alpha = 1) \Pr(\beta_0,\sigma^2) \Pr(\alpha) d \beta_0 d \sigma^2}.
\end{equation}
In particular, we integrate over uncertainty in each of the model parameters $\beta_0$, $\sigma^2$, and $\alpha$ after computing a closed-form integral over the effect size $\beta$. These BFs can be re-written using functions $h_1$ and $h_2$ (Supplementary Methods):
\begin{equation}
BF = \frac{ \int \int \int \exp^{-n \cdot h_{1} ( \beta_0, \sigma^2, \alpha)} d \beta_0 d \sigma^2 d \alpha} {\int \int e^{-n \cdot h_{2} ( \beta_0, \sigma^2, \alpha =1)} d \beta_0 d \sigma^2}
\end{equation}
where
\begin{equation}
- n \cdot h_{1} (\beta_0, \sigma^2, \alpha) = \log \left( \Pr(y |x ; \beta_0, \sigma^2, \alpha) \cdot \Pr(\sigma^2,\beta_0) \cdot \Pr(\alpha) \right),
\end{equation}
and
\begin{equation}
- n \cdot h_2(\beta_0, \sigma^2) =\log \left( \Pr(y |x ; \beta_0, \sigma^2, \alpha = 1
) \cdot \Pr(\sigma^2,\beta_0)\right).
\end{equation}

Bayes factors are computed using multivariate Laplace approximations of integrals of the form $\int e^ {- n h(t)} dt  = e^{-n h(\hat{t})} \cdot (2\pi) ^{\frac{d}{2}} \cdot |\Sigma|^{\frac{1}{2}} \cdot n^{-\frac{d}{2}}$, where $\hat{x} = argmin_xh(x)$, and $\Sigma$ is the inverse Hessian of $h$ evaluated at $\hat{x}$ (derivations in Supplementary Methods)~\cite{rue2009approximate}. As is common in Bayesian analysis of genomic studies, we report the $\log_{10}$ transformed Bayes factors~\cite{stephens2009bayesian}. 

\subsection*{FDR calibration.}
Global false discovery rate (FDR) of the log Bayes factors $\log_{10} BF(x,y)$ is quantified using permutations. To do this, we developed a permutation that preserved the mean effects but removed any variance effects. In particular, for trait $y \in \Re^n$, we computed a mean-effect-preserving transformation as follows.
We fit a linear regression model using generalized least squares
and computed residuals $r_i = y_i - \beta_{gls} x_i$ for each sample $i$. We then randomly permuted the sample indices on $r_i$, $r_{\pi(i)}$, checking that the mean effects of $r_{\pi(i)}$ versus $x_i$ are not statistically different than zero. Finally, we set $y_{\pi(i)} = \beta_{gls} x_i + r_{\pi(i)}$.

Global FDR calibration is performed after computing the unpermuted and permuted Bayes factors, $BF^{(0)} = BF(x,y)$ and $BF^{(\pi)} = BF(x,y_{\pi})$. For $d$ being a Bayes factor threshold, true positives (TP) and false positives (FP) are estimated using these BFs as $\widehat{TP} = \# \{j: |BF_j^{(0)}| >d \}$ and $\widehat{FP} + \widehat{FP}(d) = \# \{j: |BF_j^{(\pi)}| >d \}$ respectively. Thus, the estimated FDR at cut-off $d$ is computed as $\widehat{FDR}(d) = \frac{\widehat{FP}(d)} {\widehat{TP}(d)+\widehat{FP}(d)}$. For a specific FDR threshold, the calibrated threshold $d_{FDR}$ is computed from the data, and the pairs $(x,y)$ with $BF(x,y) > d_{FDR}$ are reported.

\subsection*{Comparative tests: Levene, Brown-Forsythe, and CLS.}
The Levene, Brown-Forsythe, and CLS tests were implemented and applied for comparison with BTH. The Brown-Forsythe and Levene tests both belong to the general Levene family of tests for equality of variance across $k$ subgroups~\cite{levene1961robust,brown1974small}. For $n$ samples corresponding to categorical covariate $\bx$ $\in \{1,2, \ldots k\}^n$, the trait $\by \in \Re^n$ is modeled as $y_i  \sim \N(\beta_0 + \beta x_i, \sigma^2_{x_i})$. The null hypothesis $H_0$ corresponds to equal variances across subgroups $\sigma^2_j = \sigma^2_{\ell}$ for all $j, \ell \in \{1,2,\ldots, k\}$. The alternative hypothesis $H_A$ corresponds to nonequal variances across subgroups $\sigma_j^2 \neq \sigma_{\ell}^2$ for at least one pair $j \neq \ell$, $j, \ell \in \{1,2,\ldots, k\}$. Let $n_t$ be the 
number of samples $\bx$ $\in \{1, 2, \ldots, k\}^n$ for which $x_i =  t$. Consider a partition of the entries $y_i$, $i \in \{1,2, \ldots, n\}$, into $k$ sub-vectors $\omega_{t}$, $t \in \{1,2, \ldots, k \}$, with entries $\omega_{ts}$, $s \in \{1,2,\ldots, n_t \}$,  referring to the $s$th value $y_i$ which corresponds to the subgroup $x_i = t$. The Levene family test statistic is then defined as:
\begin{equation}
W = \frac{ (n-k) \sum_{t=1}^{k} (\bar{z}_{t} - \bar{z})^2}{(k-1) \sum_{t=1}^{k} \sum_{s=1}^{n_t} ({z}_{ts} - \bar{z}_{t})^2}, 
\end{equation}
with $z_{ts} = | \omega_{ts} - \bar{\omega}_{t} |$. For a fixed $t \in \{1,2, \ldots, k\}$ $\bar{\omega}_t $ is the mean of all $\omega_{ts}$ with $s \in \{1,2, \ldots, n_t \}$. Similarly, $\bar{z}_t$ is the mean over the values $z_{ts}$, while $\bar{z}$ is the overall mean, over the entries of the vector trait $y$.   
 When each $\bar{\omega}_{t}$ is the median of the  $\omega_{ts}$ values, instead of their mean, the test is called the Brown-Forsythe test. 
 
Significance and global FDR are computed based on permutations as with BTH, replacing BFs with p-values.

The CLS test was implemented by computing residuals $r_i = y_i -
\beta_0 - \beta x_i$ where $\beta_0$ and $\beta_i$ were fit using
generalized least squares, modeling the trait $y_i$ conditional on the
genotype $x_i$ for each individual $i$ using linear regression. The
Spearman rank correlation test between the squared residuals, $r_i^2 =
(y_i - ({\hat \beta_0} + {\hat \beta_i} x_i))^2$, and the genotypes,
$x_i$, correspond to a test for variance QTLs. This implementation of
CLS followed the description in earlier work~\cite{Brown2014}.

\subsection*{Regression distribution classifier.}

We trained a random forest classifier ({\tt RandomForest} in scikit-learn~\cite{scikit-learn}, version 0.16.1) to distinguish between six possible departures from the ideal heteroskedastic model. For each distribution class (the BTH model, additive variance model, exponential mean model, exponential residue model, log Gaussian, gamma, and bimodal models; see Supplemental Methods for descriptions) 
and for four parameter configurations (scenarios $\alpha \neq 0$ and $\beta \neq 0$, $\alpha \neq 0$ and $\beta = 0$, $\alpha = 0$ and $\beta \neq 0$, and $\alpha = 0$ and $\beta = 0$), we generated $50$ samples of observed data from those six models. One-sided Kolmogorov-Smirnov statistics between each of these samples and $79$ probability density functions were computed. Thus, every sample is represented as a point in a $79$-dimensional feature space. Performance of the RF classifier was evaluated using five-fold cross validation. The generalization of the classifier was quantified using a precision recall curve (Fig.~S36). 

\subsection*{Cardiovascular and Pharmacogenetics (CAP) study data.}

Gene expression levels from $10,195$ genes in lymphoblastoid cell lines (LCLs) created from $480$ genotyped individuals were downloaded from the Gene Expression Omnibus (GSE36868). 
Genotypes for $387,514$ SNPs and eight other covariates were available through dbGaP (Study Accession phs000481.v1.p1)~\cite{mangravite2013statin}.

We processed the raw gene expression data as follows.
\begin{enumerate}
\item \emph{Log transform:} A $\log_2$ transformation was applied to each entry of the gene expression matrix~\cite{};
\item \emph{Control for latent population structure:} We computed the first two principal components $x_{PC1}$, $x_{PC2}$ of the genotype matrix via singular value decomposition (SVD).
\item \emph{Control for known covariates; mean center:} For each vector $y_j$ in matrix $\bY$, corresponding to single gene $j$ across all $n$ samples, a linear model
$y_j = \lambda_0 + \lambda_{age} \cdot x_{age} + \lambda_{sex} \cdot x_{sex} + \lambda_{batch} \cdot x_{batch} + \lambda_{PC1} \cdot x_{PC1} + \lambda_{PC2} \cdot x_{PC2}$ was fitted to account for variation in gene expression due to sample age, sex, batch, and genotype PCs, using generalized least squares. Mean-centered residuals $r_j = y_j - \hat{\lambda}_0 - \hat{\lambda}_{age} \cdot x_{age} - \hat{\lambda}_{sex} \cdot x_{sex} - \hat{\lambda}_{batch} \cdot x_{batch} - \hat{\lambda}_{PC1} \cdot x_{PC1} - \hat{\lambda}_{PC2} \cdot x_{PC2}$ were computed. Concatenating the $r_j$ vectors forms a normalized expression matrix.

\item \emph{Control for unknown covariates:} We computed the first two principal components of the normalized expression matrix using SVD. We used linear regression as in the previous step to control for the linear effects of these two PCs in the normalized gene expression matrix. 
\end{enumerate}  
The resulting matrix is the processed gene expression matrix.

\subsection*{HapMap phase 2 methylation study data.}
Processed DNA methylation data using the Infinium HumanMethylation450 BeadChip platform were downloaded from the Gene Expression Omnibus (GEO), Accession number GSE36369~\cite{heyn2013dna} on August 6, 2015. We extracted the methylation data corresponding to $280$ individuals for whom genotypes were available. Genotypes spanning $166,947$ common single nucleotide polymorphisms (SNPs) were obtained from genome-wide DNA array Human Variation Panel studies ~\cite{niu2010radiation,kalari2010copy} through accession numbers GSE24260 and GSE24274, which used Illumina 550K and Illumina 650K array data respectively. 
We filter poor quality CpG probes by removing those methylation sites where 90\% of the samples at that site are hypo or hyper-methylated, less than 2\% or greater than 98\% methylated, respectively. From a total of $54750$ total CpG probes, we filter $2112$ probes to end up with $52638$ probes to test for association with genotypes.

\section*{Acknowledgments}
BD and BEE were funded by NIH R00 HG006265 and NIH R01 MH101822. 
 All data are publicly available. The CAP gene expression data were acquired through GEO GSE36868, and the genotype data were acquired through dbGaP, acquisition number phs000481, and generated from the Krauss Lab at the Children's Hospital Oakland Research Institute. This work was supported in part by U19 HL069757: Pharmacogenomics and Risk of Cardiovascular Disease.  We acknowledge the PARC investigators and research team, supported by NHLBI, for collection of data from the Cholesterol and Pharmacogenetics clinical trial. All code is freely available for
   use at the Engelhardt Group Website: {\tt http://beehive.cs.princeton.edu/software.html}. 
Correspondence and requests for materials
should be addressed to BEE~(email: bee@princeton.edu).

\bibliography{refs}

\end{document}